\documentclass[twocolumn,tighten,twocolappendix]{aastex63}
\pdfoutput=1 
\usepackage{amsmath,amstext}
\usepackage[T1]{fontenc}
\usepackage{ae,aecompl}
\usepackage[utf8]{inputenc}
\usepackage{apjfonts} 
\usepackage[figure,figure*]{hypcap}
\usepackage{enumitem}
\usepackage{bm}
\usepackage{pifont}
\input{hyperlink-year-only-natbib-patch}

\newcommand*{\https}[1]{\href{https://#1}{#1}}

\graphicspath{{./}{}}

\shorttitle{SIDM, WDM, and Baryonic Effects on MW Subhalos}
\shortauthors{Nadler et al.}

\begin{document}

\title{The Effects of Dark Matter and Baryonic Physics on the Milky Way Subhalo Population in the Presence of the Large Magellanic Cloud}

\correspondingauthor{Ethan~O.~Nadler}
\email{enadler@carnegiescience.edu}
\author[0000-0002-1182-3825]{Ethan~O.~Nadler}
\affiliation{Carnegie Observatories, 813 Santa Barbara Street, Pasadena, CA 91101, USA}
\affiliation{Department of Physics $\&$ Astronomy, University of Southern California, Los Angeles, CA, 90007, USA}
\author[0000-0002-5209-1173]{Arka~Banerjee}
\affiliation{Fermi National Accelerator Laboratory, Cosmic Physics Center, Batavia, IL 60510, USA}
\author[0000-0002-0298-4432]{Susmita~Adhikari}
\affiliation{Department of Astronomy and Astrophysics, University of Chicago, Chicago, IL 60637, USA}
\affiliation{Kavli Institute for Cosmological Physics, University of Chicago, Chicago, IL 60637, USA}
\author[0000-0002-1200-0820]{Yao-Yuan~Mao}
\altaffiliation{NHFP Einstein Fellow}
\affiliation{Department of Physics and Astronomy, Rutgers, The State University of New Jersey, Piscataway, NJ 08854, USA}
\author[0000-0003-2229-011X]{Risa~H.~Wechsler}
\affiliation{Kavli Institute for Particle Astrophysics and Cosmology and Department of Physics, Stanford University, Stanford, CA 94305, USA}
\affiliation{SLAC National Accelerator Laboratory, Menlo Park, CA 94025, USA}

\begin{abstract}
Given recent developments in our understanding of the Large Magellanic Cloud's (LMC) impact on the Milky Way's (MW) dark matter subhalo population, we compare the signatures of dark matter and baryonic physics on subhalos in MW systems with realistic LMC analogs. In particular, we study the effects of self-interacting dark matter~(SIDM), warm dark matter (WDM), and the Galactic disk on the peak maximum circular velocity ($V_{\mathrm{peak}}$) function, radial distribution, and spatial distribution of MW and LMC-associated subhalos using cosmological dark matter-only zoom-in simulations of MW+LMC systems. For a fixed abundance of subhalos expected to host dwarf galaxies ($V_{\mathrm{peak}}\gtrsim 20\ \mathrm{km\ s}^{-1}$), SIDM and WDM can produce a similar mass-dependent suppression of the subhalo $V_{\mathrm{peak}}$ function, while disk disruption is mass independent. Subhalos in the inner regions of the MW are preferentially disrupted by both self-interactions and the disk, while suppression in WDM is radially independent. The relative abundance of LMC-associated subhalos is not strongly affected by disk disruption or WDM, but is significantly suppressed in SIDM due to self-interactions with the LMC at early times and with the MW during LMC infall at late times, erasing spatial anisotropy in the MW subhalo population. These results provide avenues to distinguish dark matter and baryonic physics by combining properties of the MW and LMC subhalo populations probed by upcoming observations of satellite galaxies and stellar streams.
\end{abstract}

\keywords{\href{http://astrothesaurus.org/uat/353}{Dark matter (353)}; \href{http://astrothesaurus.org/uat/1049}{Milky Way dark matter halo (1049)}; \href{http://astrothesaurus.org/uat/574}{Galaxy abundances (574)}}

\section{Introduction}
\label{sec:intro}

Dark matter subhalos orbiting the Milky Way (MW) provide a powerful means to test structure formation and the cold dark matter (CDM) paradigm on the smallest observationally accessible cosmic scales. For decades, the ``missing satellites problem'' concerning MW satellite galaxies \citep{Klypin9901240,Moore9907411} inspired dark matter model-building efforts focused on reducing the abundance of low-mass subhalos. This can be achieved by suppressing the primordial matter power spectrum---as in models of warm \citep{Bode0010389,Adhikari160204816}, interacting \citep{Boehm0112522,Nadler190410000}, and fuzzy \citep{Hu008506,Hui179504} dark matter---or dynamically at late times, as in models of self-interacting dark matter (SIDM; \citealt{Spergel9909386,Tulin170502358}).

Because of its size (with a mass of $\sim 10^{11}\ M_{\mathrm{\odot}}$; \citealt{Erkal181208192}) and recent infall into the MW (with a current Galactocentric distance of $\sim 50\ \mathrm{kpc}$;  \citealt{Kallivayalil13010832}), the Large Magellanic Cloud (LMC) system significantly affects the MW's total subhalo population. For example, the LMC is accompanied by its own ultra-faint dwarf galaxies~\citep{Kallivayalil180501448,Patel200101746}, yielding an overabundance of subhalos and satellites in its vicinity~\citep{PaperII}. Thus, it is crucial to include realistic LMC analogs---constrained in terms of their mass, orbital properties, and associated satellites---in predictions for the MW subhalo population, including in alternative dark matter models.

Here, we investigate subhalo populations in cosmological zoom-in simulations of two MW-like systems, both of which include realistic LMC analogs and are consistent with the population of observable MW satellite galaxies, in popular non-CDM models and in the presence of baryons. In particular, we present new SIDM resimulations of these systems and we compare the resulting subhalo populations to those expected in warm dark matter (WDM) and in the presence of a galactic disk. We study the impact of SIDM, WDM, and baryons separately to isolate the mechanisms that influence the formation and survival of subhalos in each scenario.

We find that the peak maximum circular velocity ($V_{\mathrm{peak}}$) function, radial distribution, and spatial distribution of the total MW subhalo population respond differently to SIDM, WDM, and baryonic physics. Strikingly, a significantly larger fraction of LMC-associated subhalos are disrupted in our SIDM simulations than in our WDM or Disk models, erasing the expected spatial anisotropy in the MW subhalo population introduced by the LMC system. This extra disruption is caused by self-interactions between LMC-associated subhalos and the LMC at early times and self-interactions with the MW halo during the LMC system's radial, high-velocity infall into the MW at late times. These mechanisms reveal a unique signature of SIDM physics on the total MW subhalo population that depends on both the history and present configuration of the MW--LMC system.

Although many different dark matter models can impact subhalos similarly (e.g., \citealt{Buckley171206615}), our results demonstrate that the total MW subhalo population can be used to disentangle the imprints of new dark matter physics from each other and from baryonic effects when accounting for the impact of the LMC. Excitingly, forthcoming surveys including the Vera C.\ Rubin Observatory Legacy Survey of Space and Time (LSST) will enable unprecedented measurements of MW and LMC-associated satellite galaxies and stellar streams, which trace the underlying subhalo population (e.g., \citealt{Drlica-Wagner190201055}).





This Letter is organized as follows. In Section \ref{sec:methods}, we describe our MW-like CDM simulations, our new SIDM resimulations of these systems, and our WDM and disk disruption models. We describe how each scenario impacts the MW and LMC subhalo populations in Section \ref{sec:results}. We discuss our results and conclude in Section \ref{sec:discussion}. Throughout, we define virial quantities using the \cite{Bryan9710107} critical overdensity~$\Delta_{\rm vir}\simeq 99.2$ corresponding to the cosmological parameters adopted in our simulations: $h = 0.7$,~$\Omega_{\rm m} = 0.286$, $\Omega_{\rm b} = 0.047$, and $\Omega_{\Lambda} = 0.714$ \citep{Mao150302637}.

\begin{figure*}[t]
\includegraphics[scale=0.3]{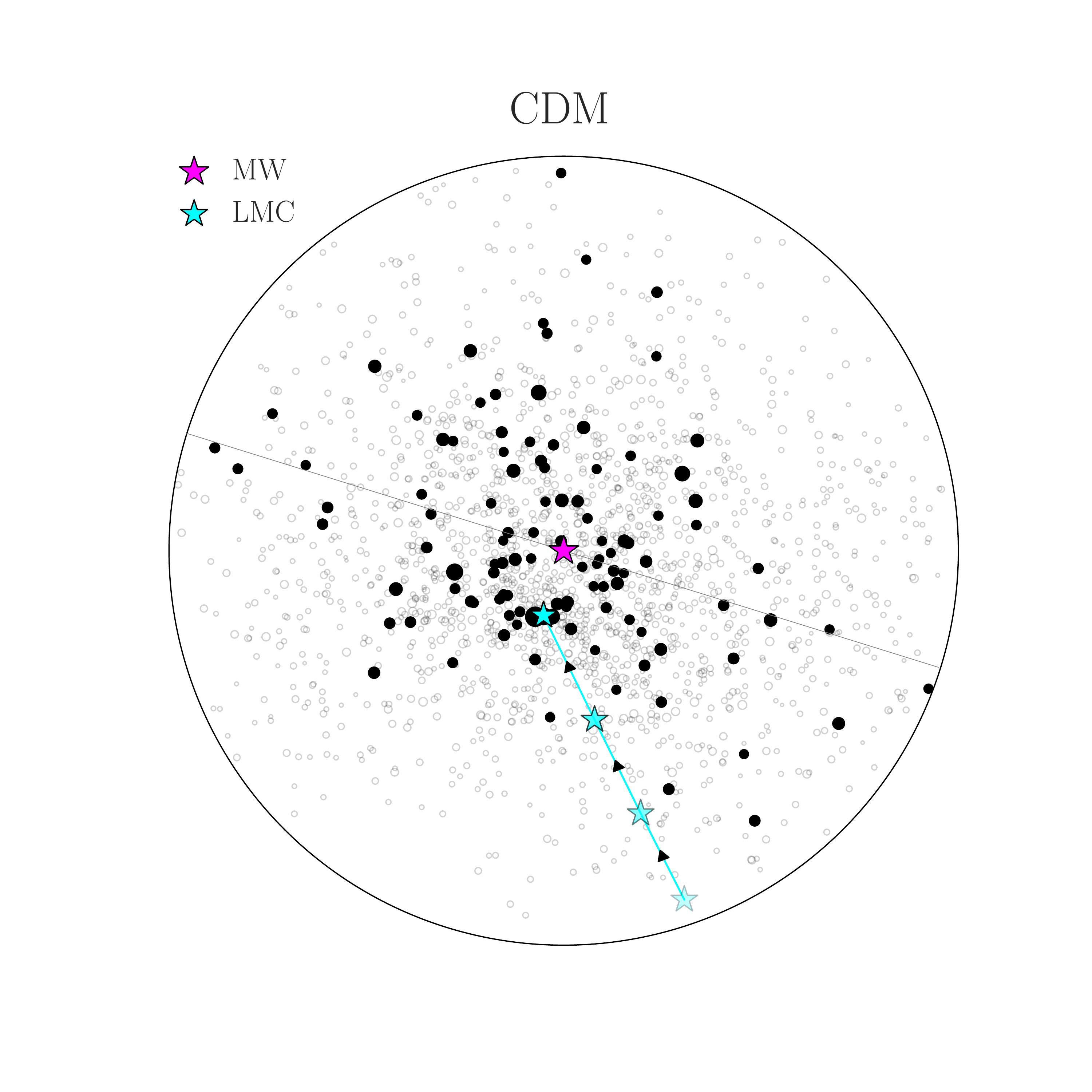}
\hspace{-3mm}
\includegraphics[scale=0.3]{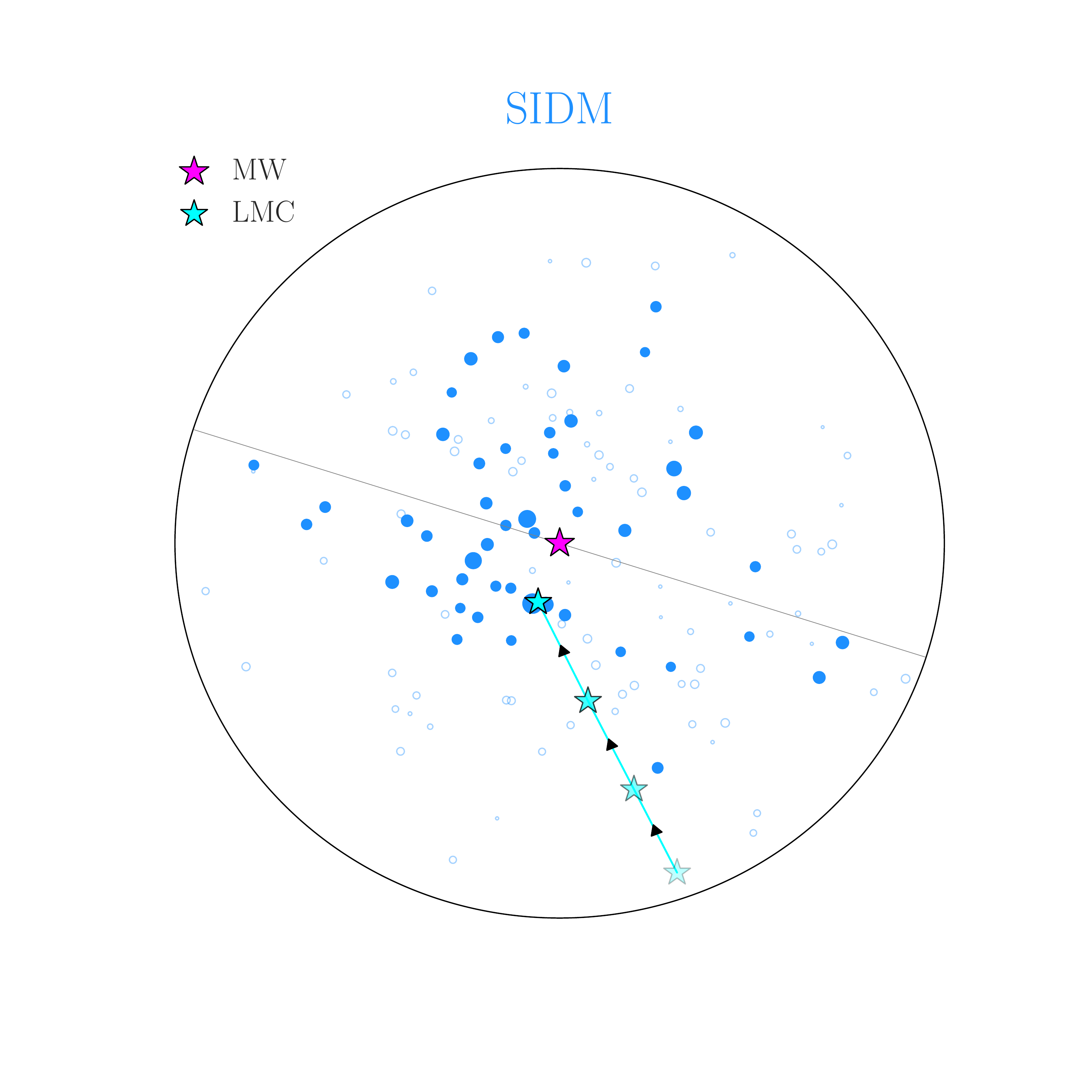}
\\
\includegraphics[trim={0 3cm 0 3cm},scale=0.3]{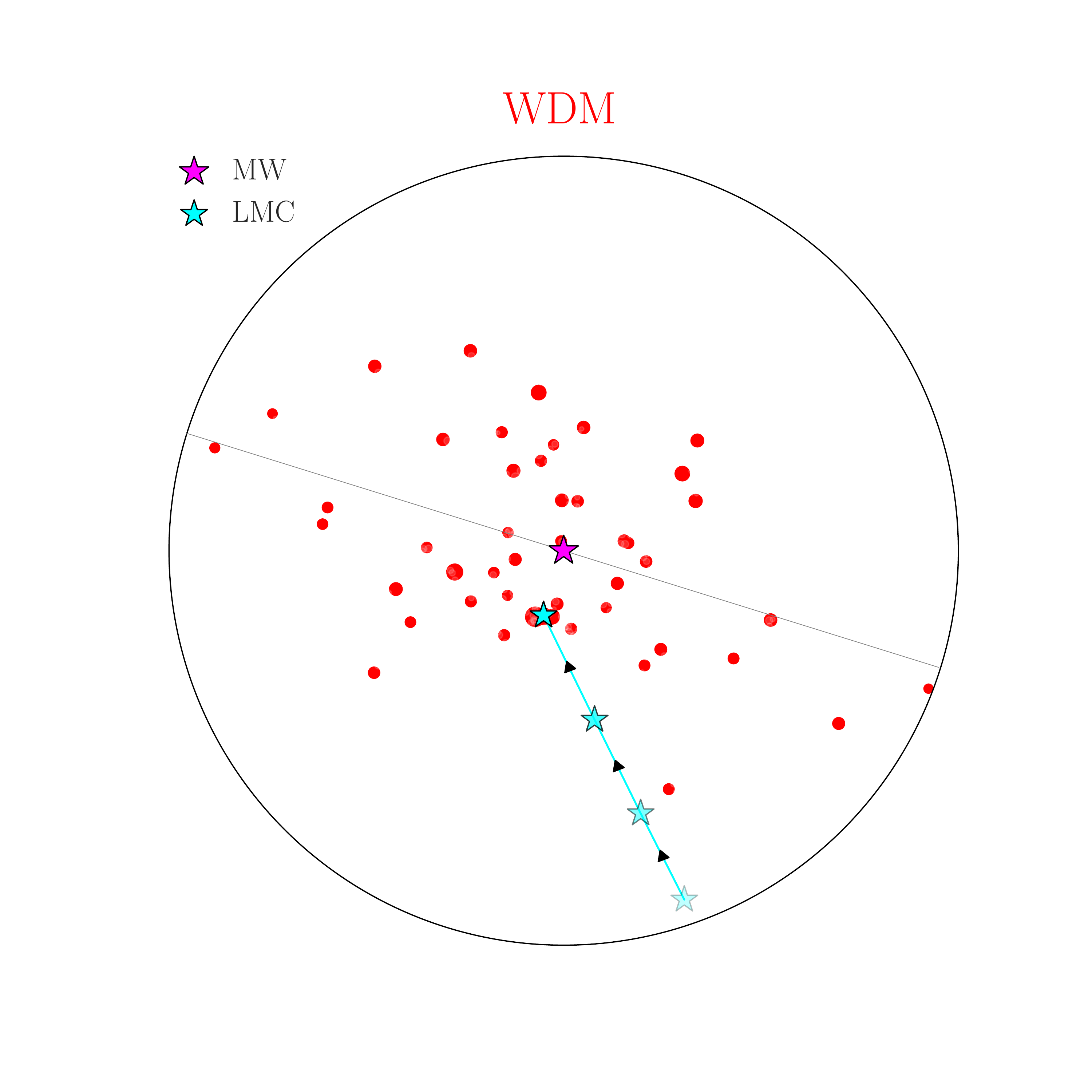}
\hspace{-3mm}
\includegraphics[trim={0 3cm 0 5cm},scale=0.3]{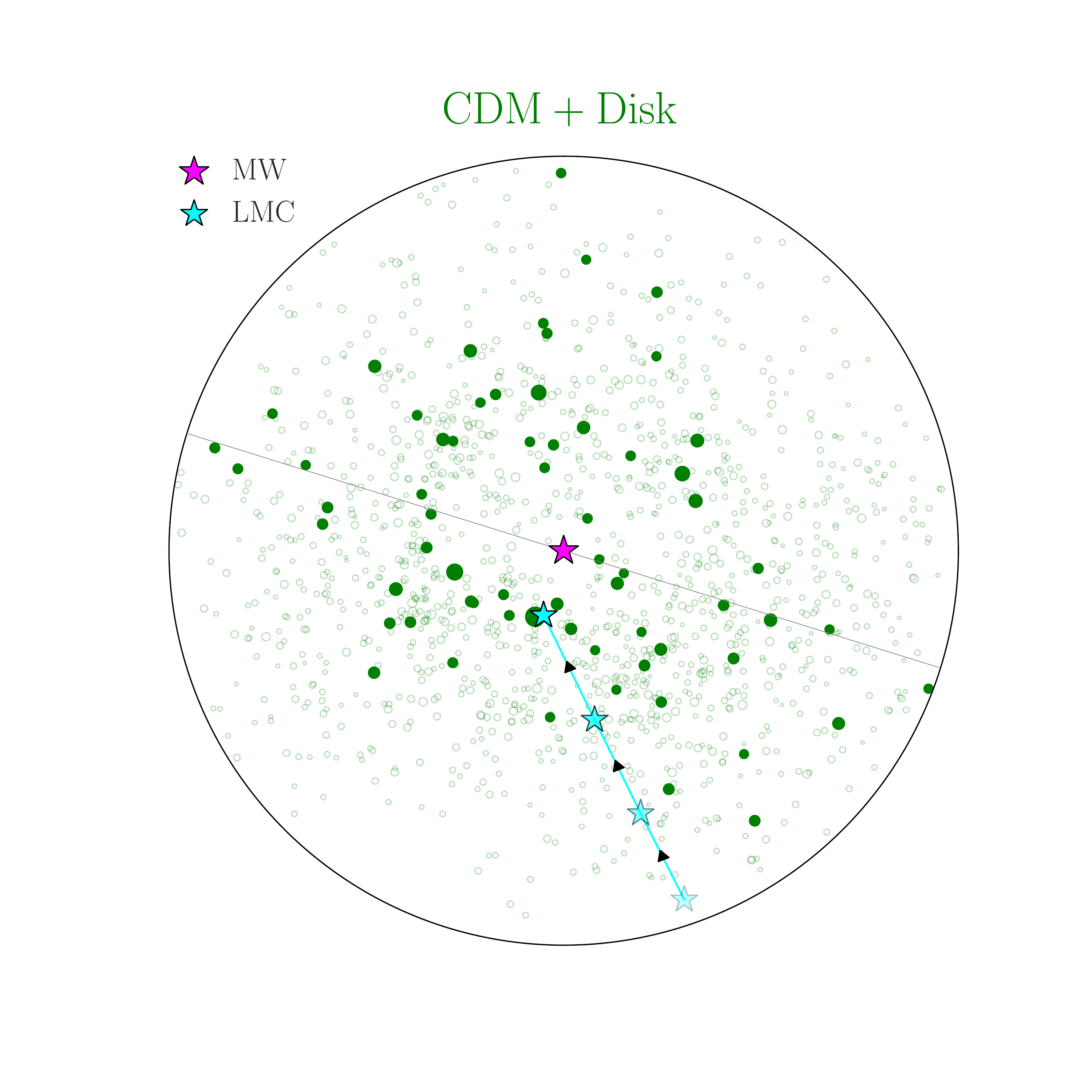}
\caption{Projections of the subhalo population in one of our MW-like simulations in CDM (top left), in our SIDM resimulation of this system (top right), and in our WDM (bottom left) and disk disruption (bottom right) models applied to the CDM simulation. In the WDM and disk disruption cases, we apply a cut on survival probability to match the abundance of subhalos with $V_{\mathrm{peak}}>20\ \mathrm{km\ s}^{-1}$ in our SIDM resimulation. Filled markers show subhalos with $V_{\mathrm{peak}}>20\ \mathrm{km\ s}^{-1}$, unfilled markers show subhalos below this $V_{\mathrm{peak}}$ threshold, and marker size is proportional to $V_{\mathrm{peak}}$. In each panel, the black circle shows the virial radius of the MW host halo ($\sim 300\ \mathrm{kpc}$), the gray line shows the plane defined by the position of the MW (magenta star) and LMC (cyan star) at $z=0$, and transparent cyan stars show the LMC's trajectory as it falls into the MW.}
\label{fig:vis}
\end{figure*}

\section{Methods}
\label{sec:methods}

\subsection{MW-like CDM Simulations}
\label{sec:cdm}

Our analysis is based on two CDM dark-matter-only zoom-in simulations of MW-like halos originally presented in \citet{Mao150302637}. The highest-resolution particles in these simulations have a mass of~$3\times 10^{5}\ \mathrm{M}_{\rm \odot}\ h^{-1}$, and the Plummer-equivalent softening length in the highest-resolution regions is $170\ \text{pc}\ h^{-1}$. Subhalos are well resolved down to a peak maximum circular velocity of $V_{\rm peak} \approx 10\ \rm{km\ s}^{-1}$, where $V_{\rm{peak}}$ is the largest maximum circular velocity a halo attains over its entire history \citep{Mao150302637}. Halo catalogs and merger trees were generated using the {\sc Rockstar} halo finder and the {\sc consistent-trees} merger tree code \citep{Behroozi11104372,Behroozi11104370}.

These simulations include both Gaia-Enceladus-like mergers at early times and realistic LMC analog systems with current Galactocentric distances of $\sim 50\ \mathrm{kpc}$ that fall into the MW within the last~$\sim 2\ \mathrm{Gyr}$ on orbits consistent with LMC proper motion measurements \citep{Kallivayalil13010832}. The properties of the MW host halos, LMC analogs, and Gaia-Enceladus analogs are described in \citet[Section 3.2 and Appendix A.3]{PaperII}. When associated with satellite galaxies, the subhalo populations in these simulations  are consistent with current Dark Energy Survey (DES; \citealt{1601.00329}) and Pan-STARRS1 (PS1; \citealt{1612.05560}) measurements of the MW satellite population, which probe subhalos with peak virial masses down to~$\sim 10^8\ M_{\mathrm{\odot}}$ 
\citep[Section 7.1]{PaperII}.

As visualized in Figure \ref{fig:vis}, these CDM simulations feature an overabundance of subhalos in the vicinity of the LMC, which is consistent with the population of LMC-associated satellites inferred from Gaia \citep{Gaia180409381} proper motion measurements and detected in DES and PS1 \citep[Section 7.2]{PaperII}.

\subsection{SIDM Simulations}
\label{sec:sidm}

We resimulate both MW-like systems described above in the presence of dark matter self-interactions using the modified version of {\sc GADGET-2} presented in \cite{Banerjee190612026}. 
We only study subhalos with~$V_{\rm peak} > 20\ \rm{km\ s}^{-1}$, which are resolved with $\gtrsim 1000$ particles at the time $V_{\mathrm{peak}}$ is achieved, to avoid difficulties introduced by halo finder incompleteness; with this choice, the total abundance of surviving plus disrupted subhalos differs at the percent level relative to CDM for the SIDM model considered in this study (\citealt{Nadler200108754}, Appendix B). Furthermore, this $V_{\rm peak}$ threshold roughly corresponds to the minimum halo mass associated with observed MW satellite galaxies based on abundance matching \citep{PaperII}. Our results therefore directly inform the interpretation of MW satellite galaxy population measurements.

\cite{Nadler200108754} reported significant levels of subhalo disruption in SIDM simulations of a MW-mass halo for SIDM models with cross sections of $\sim 1\ \mathrm{cm}^{2}\ \mathrm{g}^{-1}$. The efficiency of subhalo disruption in SIDM models that do not yield core collapse is most sensitive to the self-interaction cross section at the orbital velocity scale of $\sim 200\ \mathrm{km\ s}^{-1}$ set by the MW host halo \citep{Nadler200108754}. We therefore simulate an SIDM model with a total cross section of $0.1\ \mathrm{cm}^{2}\ \mathrm{g}^{-1}$ that is velocity independent on the scales of interest for our study, and we expect our results to generalize to velocity-dependent models that avoid core collapse in low-mass subhalos. Because there may be significant differences between the efficiency of subhalo disruption predicted by various SIDM $N$-body codes (e.g., \citealt{Nadler200108754}), we only focus on the mass, radial, and spatial dependence of the trends imprinted by this particular SIDM model on the MW and LMC subhalo populations, and we refer the reader to \cite{Nadler200108754} for a study of how these effects scale with cross-section amplitude.

The top-right panel of Figure \ref{fig:vis} shows the subhalo population in one of our SIDM resimulations, which is significantly altered relative to CDM; we compare these populations statistically in Section~\ref{sec:results}.

\subsection{WDM Model}
\label{sec:wdm}

Thermal relic WDM is a benchmark model that can be mapped to a wide range of dark matter particle scenarios with suppressed small-scale structure and that has been simulated extensively. To model its impact on the subhalo populations in our MW-like simulations, we assign a weight to each subhalo in our CDM simulations based on the probability it survives in a corresponding WDM simulation. We estimate this probability using a fitting function that describes the suppression of the WDM subhalo mass function in simulations of MW-mass systems. Several fitting functions have been reported based on different WDM simulations \citep{Schneider11120330,Anderhalden12122967,Angulo13042406,Lovell13081399,Bose160407409}. Although these WDM simulation results vary below the cutoff scale, they are in excellent agreement near and above the WDM half-mode mass (e.g., see \citealt{Angulo13042406}); the subhalos we study are almost exclusively in this regime for the fiducial WDM particle mass derived below.

We express the suppression of the WDM subhalo mass function relative to that in CDM as
\begin{equation}
\frac{dN_{\rm{WDM}}}{dM} \equiv f_{\mathrm{WDM}}\left(M, M_{\rm{hm}}\right) \frac{dN_{\rm{CDM}}}{dM},
\end{equation}
where $dN_{\rm{WDM}}/dM$ ($dN_{\rm{CDM}}/dM$) is the WDM (CDM) subhalo mass function. Here, $f_{\mathrm{WDM}}$ is a suppression factor that depends on peak subhalo mass virial mass $M$ and WDM particle mass $m_{\mathrm{WDM}}$ via the half-mode mass \citep{PaperIII}
\begin{equation}
M_{\mathrm{hm}}(m_{\mathrm{WDM}}) = 5 \times 10^8 \left(\frac{m_{\mathrm{WDM}}}{3\  \mathrm{keV}}\right)^{-10/3} M_{\mathrm{\odot}}.
\label{eq:mhm_mwdm}
\end{equation}
To enable comparisons with previous WDM constraints from MW satellite galaxies (e.g., \citealt{PaperIII}), we adopt the \cite{Lovell13081399} suppression
    \begin{equation}
    f_{\mathrm{WDM}}(M,m_{\mathrm{WDM}}) = \left[1+\left(\frac{\alpha M_{\mathrm{hm}}(m_{\mathrm{WDM}})}{M}\right)^{\beta}\right]^{\gamma},\label{eq:wdm_shmf}
\end{equation}
where $\alpha=2.7$, $\beta=1.0$, and $\gamma=-0.99$. 

When evaluating WDM subhalo population statistics, we assign each subhalo a weight equal to $f_{\mathrm{WDM}}$. In order to easily interpret comparisons between our SIDM and WDM results, we choose $m_{\mathrm{WDM}}$ such that the abundance of subhalos with $V_{\mathrm{peak}}>20\ \mathrm{km\ s}^{-1}$ predicted by Equation \ref{eq:wdm_shmf} matches that in our SIDM resimulations above the same threshold. This yields $m_{\mathrm{WDM}}=2.8\ \mathrm{keV}$. This WDM particle mass is ruled out by a variety of small-scale structure probes (e.g., \citealt{Viel1308804,Irsic179602,Gilman190806983,Hsueh190504182,Newton201108865,PaperIII}), but we emphasize that there are large theoretical uncertainties underlying the detailed relation between SIDM cross section and subhalo disruption \citep{Nadler200108754}. We therefore leave a detailed investigation of SIDM constraints to future work.

This procedure assumes that the impact of WDM on the MW subhalo population can be expressed purely as a function of peak subhalo mass. Thus, our WDM model does not explicitly alter additional properties of the MW subhalo population, including its radial distribution (e.g., see \citealt{Lovell210403322}). Subhalos of different masses above our $V_{\mathrm{peak}}$ threshold do not exhibit significantly different radial distributions in our CDM simulations; thus, our predicted WDM and CDM subhalo populations have similar radial profiles, consistent with the results of WDM simulations for the subhalo masses considered in this study (e.g., \citealt{Anderhalden12122967,Lovell13081399,Bose160407409,Lovell210403322}).

The bottom-left panel of Figure \ref{fig:vis} qualitatively shows that WDM suppresses the abundance of low-mass subhalos relative to CDM, by construction according to Equation \ref{eq:wdm_shmf}, without significantly affecting their spatial distribution.

\subsection{Disk Disruption Model}
\label{sec:disk}

Baryons mainly affect the abundance and spatial distribution of subhalos in MW-mass systems through tidal disruption due to the central galaxy, which most strongly impacts subhalos with close pericentric passages (e.g., \citealt{D'Onghia09073482,Garrison-Kimmel170103792,Nadler171204467,Richings181112437,Webb200606695}). In analogy to our WDM procedure, we model the impact of baryons on the MW subhalo population by assigning a weight to each subhalo based on the probability that it would survive in a corresponding hydrodynamic simulation. This approach is convenient because it can flexibly accommodate predictions from different hydrodynamic simulations without the need to resimulate each MW-like system of interest.

Here, we adopt the fitting function from \cite{Samuel190411508}, which is derived using a suite of zoom-in simulations of MW-mass halos from the Feedback in Realistic Environments (FIRE) project and is consistent with the disruption efficiency inferred from MW satellite observations \citep{PaperII}.\footnote{\href{https://fire.northwestern.edu/}{https://fire.northwestern.edu/}} \cite{Samuel190411508} find that subhalo disruption in these simulations can be modeled as a function of subhalo Galactocentric radius,
\begin{equation}
\frac{dN_{\rm{disk}}}{dM} \equiv f_{\mathrm{disk}}\left(r\right) \frac{dN_{\rm{CDM}}}{dM},\label{eq:disk_shmf}
\end{equation}
where
\begin{equation}
f_{\mathrm{disk}}(r) = \begin{cases}
      0, & 0\leq r < r_0, \\
      a\left[1-\exp\left( -\frac{r-r_0}{r_1}\right) \right], & r \geq r_0.
    \end{cases}\label{eq:fdisk}
\end{equation}
We use the best-fit parameters from \cite{Samuel190411508} for the differential suppression of subhalos with peak virial mass $M>8\times 10^8\ M_{\mathrm{\odot}}$, which are most relevant for our results. This corresponds to $a=0.8$, $r_0=8\ \mathrm{kpc}$, and $r_1=78\ \mathrm{kpc}$ in Equation \ref{eq:fdisk}. Following our WDM treatment, we assign each subhalo a weight of $f_{\mathrm{disk}}$ when evaluating subhalo population summary statistics. 

Several other algorithms and fitting functions have been proposed to model the impact of baryons on subhalo populations in MW-mass systems. The prescription above yields subhalo populations that agree reasonably well with those predicted by the \cite{Nadler171204467} model, with slightly more severe disruption at small Galactocentric radii. These predictions are also consistent with the amplitude and radial dependence of subhalo disruption predicted by other embedded disk potential (e.g., \citealt{Garrison-Kimmel170103792,Kelley181112413,Robles190301469}) and hydrodynamic (e.g., \citealt{Zhu150605537,Garrison-Kimmel180604143,Richings181112437,Stafford200403872}) simulations. Importantly, none of these studies find that the efficiency of disk disruption is a strong function of subhalo mass for the range of peak virial masses relevant to our work. Thus, there is no explicit mass dependence in Equation \ref{eq:fdisk}.

The bottom-right panel of Figure \ref{fig:vis} illustrates the suppression of subhalo abundances in the inner regions of the MW caused by the Galactic disk.

\section{Results}
\label{sec:results}

We now present our results, focusing on the suppression of subhalo abundance as a function of $V_{\mathrm{peak}}$ (Section \ref{sec:shmf}), radial distance (Section \ref{sec:radial}), and LMC association (Section \ref{sec:spatial}) as summarized in Figure \ref{fig:vpeak_function} and Table \ref{tab:sidm_sims}. 

Our SIDM results follow directly from resimulations of our two MW-like halos, while our WDM and Disk results convolve the fitting functions in Equations \ref{eq:wdm_shmf} and \ref{eq:fdisk} with the properties of the underlying MW+LMC subhalo populations in our CDM simulations. We note that the $V_{\mathrm{peak}}$, radial, and spatial distributions of subhalos are weakly correlated in typical MW-mass halos \citep{Springel08090898}. Although the LMC system can enhance these correlations in principle, we find empirically that they remain weak in our MW-like simulations. Thus, it follows that our WDM model mainly impacts the shape of the $V_{\mathrm{peak}}$ function without strongly affecting the shape of the radial distribution, while the Disk model mainly suppresses the abundance of subhalos in the inner regions of the MW without strongly affecting the shape of the $V_{\mathrm{peak}}$ function.

\subsection{Subhalo $V_{\mathrm{peak}}$ Function}
\label{sec:shmf}

\begin{figure*}[t!]
\hspace{-1mm}
\includegraphics[scale=0.35]{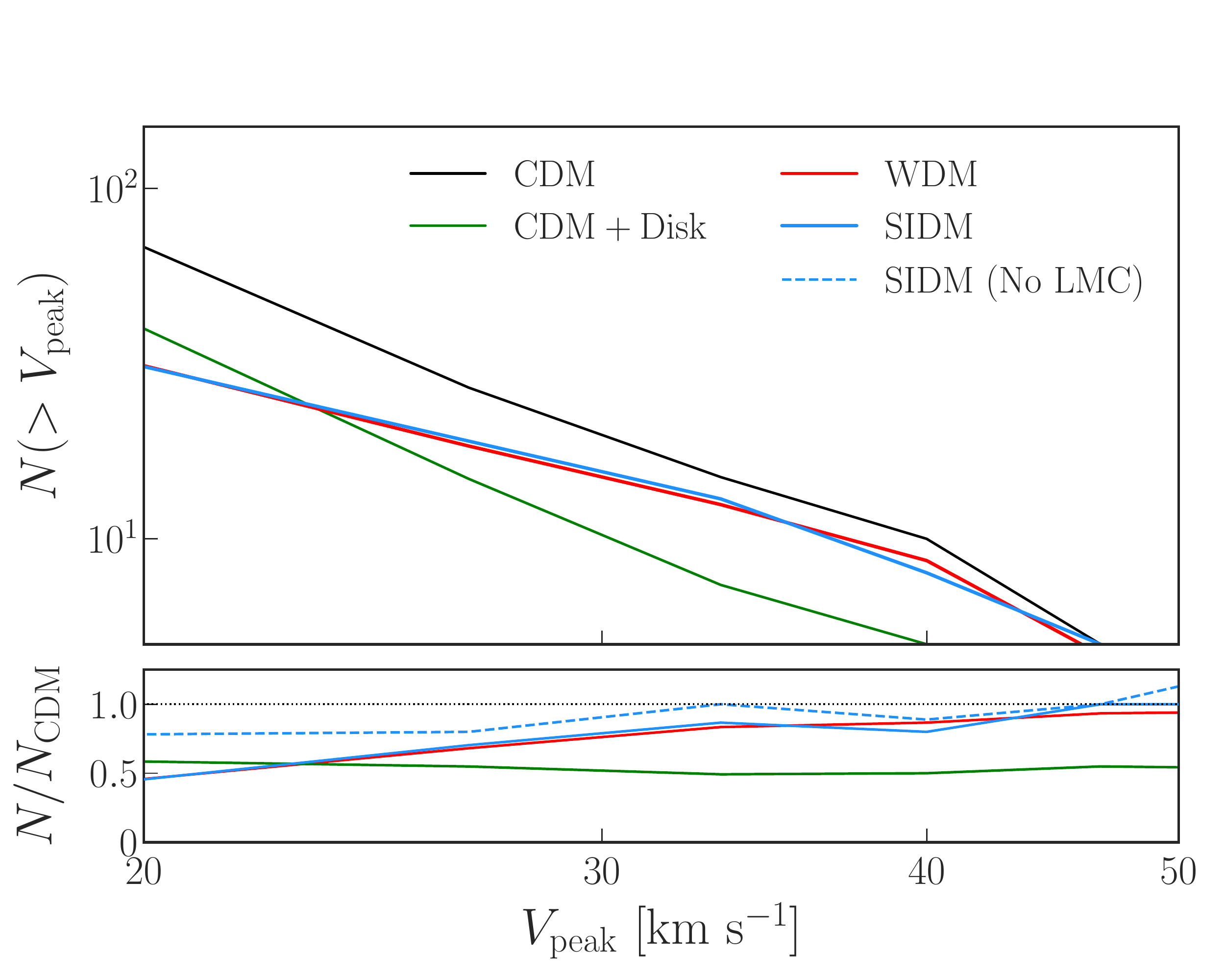}
\hspace{1.5mm}
\includegraphics[scale=0.35]{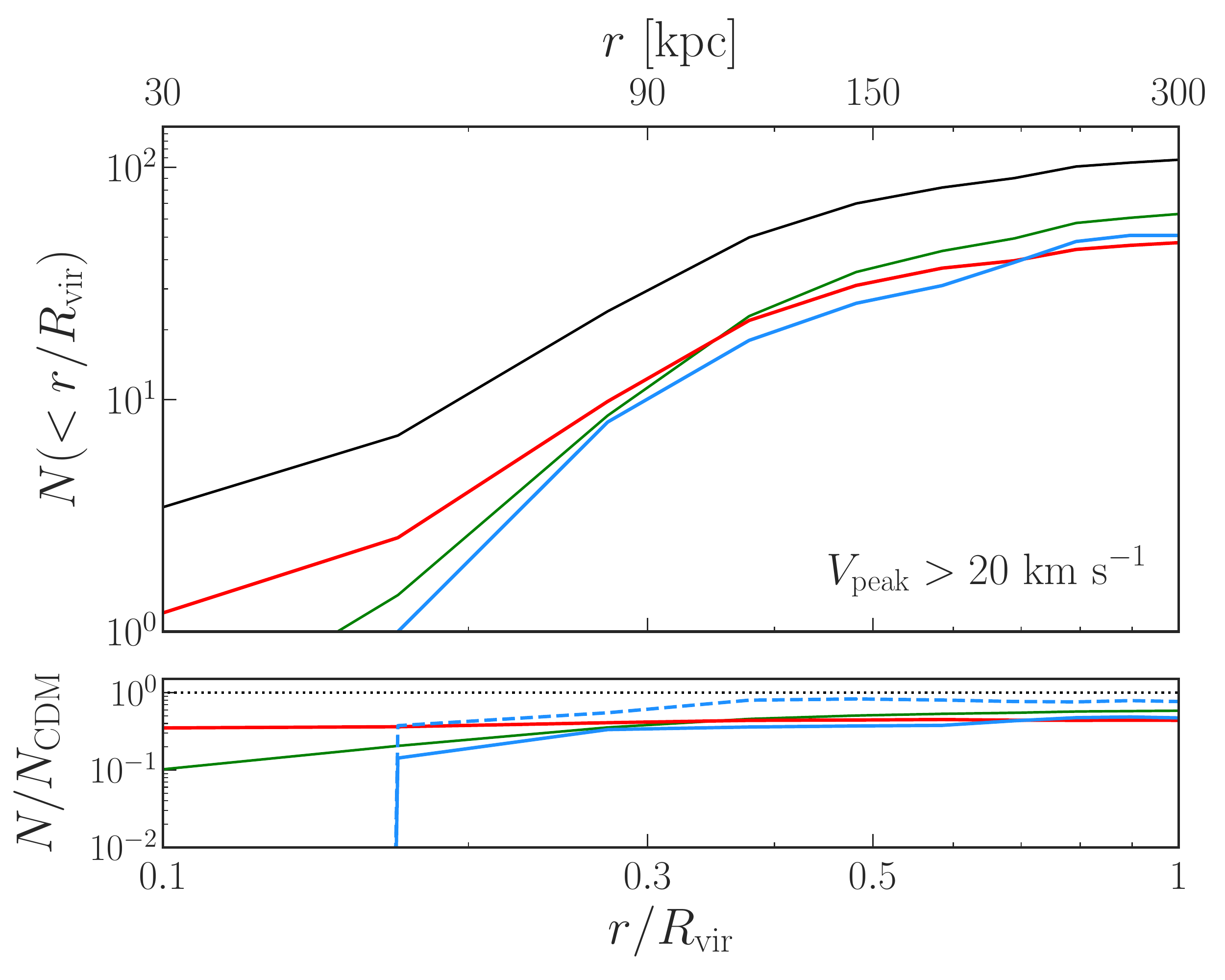}
\caption{The effects of SIDM, WDM, and the Galactic disk on the $V_{\mathrm{peak}}$ function and radial distribution of surviving subhalos in one of our MW-like simulations. Left panel: number of subhalos as a function of peak maximum circular velocity $V_{\mathrm{peak}}$ in CDM (black), SIDM (blue), and as predicted by applying our WDM (red) and disk disruption (green) models to the CDM simulation. Disk disruption is approximately $V_{\mathrm{peak}}$ independent, while SIDM and WDM preferentially disrupt lower-$V_{\mathrm{peak}}$ subhalos. Right panel: same as the left panel, but for the cumulative radial distribution of subhalos in units of the MW host halo virial radius. The suppression of subhalo abundance in WDM is not a strong function of Galactocentric radius, while SIDM and the disk preferentially disrupt subhalos at small radii. In the bottom panels, the dashed blue lines show the suppression of the subhalo population for the same SIDM model in an MW-mass system without a realistic LMC analog. Note that the WDM and Disk results are weighted by subhalo survival probability and can therefore have $N/N_{\mathrm{CDM}}>0$ in regions where $N(<r/R_{\mathrm{vir}})<1$.}
\label{fig:vpeak_function}
\end{figure*}

The left panel of Figure \ref{fig:vpeak_function} shows the cumulative number of surviving subhalos as a function of $V_{\mathrm{peak}}$ in each model.\footnote{Our results are qualitatively similar when expressed in terms of subhalo mass functions (e.g., using peak virial mass) rather than $V_{\mathrm{peak}}$ functions.} Disk disruption results in a roughly mass-independent suppression of subhalo abundance relative to CDM. This is due to the fact that subhalos in our disk disruption model are only suppressed based on their Galactocentric radius, which is not significantly correlated with subhalo mass in our CDM simulations. The subhalo $V_{\mathrm{peak}}$ function in our disk disruption model therefore retains an approximate power-law form with a lower normalization than in CDM.

Meanwhile, WDM and SIDM both preferentially reduce the abundance of low-mass subhalos relative to CDM. In WDM, this results from free-streaming, which suppresses the linear matter power spectrum on small scales and inhibits the formation of low-mass halos (e.g., \citealt{Schneider13030839}). In SIDM, this mass-dependent suppression is caused by enhanced tidal stripping in the presence of ram-pressure interactions between subhalos and the MW host halo, as described in \cite{Nadler200108754}. Thus, unlike for WDM, the reduction in subhalo abundance in our SIDM simulations does not necessarily imply a corresponding reduction in the abundance of satellite galaxies, which may survive even after their subhalos are heavily stripped.

While we match the amplitude of WDM and SIDM subhalo abundances by construction, the detailed agreement between the corresponding $V_{\mathrm{peak}}$ functions shown in Figure \ref{fig:vpeak_function} is coincidental. In particular, these $V_{\mathrm{peak}}$ functions deviate for $V_{\mathrm{peak}}<20\ \mathrm{km\ s}^{-1}$ or for a different choice of the $V_{\mathrm{peak}}$ threshold at which subhalo abundances are matched. This is expected because low-mass subhalos are \emph{dynamically} disrupted in SIDM, while their \emph{formation} is suppressed in WDM due to the effects of free-streaming on the linear matter power spectrum. Nonetheless, our qualitative result---i.e., that the abundance of lower-mass subhalos is preferentially suppressed compared to higher-mass subhalos by WDM or SIDM and not by the disk---remains robust.

Interestingly, the suppression of the subhalo $V_{\mathrm{peak}}$ function for the same SIDM model run on the MW-mass system without a realistic LMC analog from \cite{Nadler200108754} is less severe in terms of both amplitude and mass dependence compared to the MW-like systems we study here (see the dashed blue line in the bottom panel of Figure \ref{fig:vpeak_function}).\footnote{We do not compare the absolute number of subhalos in this case because the properties of the ``No LMC'' host and our MW-like halos differ in detail.} The difference between this alternate simulation and our fiducial results is consistent with the enhanced disruption of LMC subhalos in our MW-like simulations discussed in Section \ref{sec:spatial}.

Thus, the Galactic disk suppresses the abundance of subhalos in MW-like systems relative to CDM in a mass-independent manner, while the abundance of lower-mass subhalos is preferentially suppressed in WDM and SIDM.

\subsection{Subhalo Radial Distribution}
\label{sec:radial}

The right panel of Figure \ref{fig:vpeak_function} shows the radial distribution of surviving subhalos in units of the MW host halo virial radius. We observe the following trends:
\begin{enumerate}
    \item Zero subhalos with $V_{\mathrm{peak}}>20\ \mathrm{km\ s}^{-1}$ survive within a Galactocentric distance of $\sim 30\ \mathrm{kpc}$ in our fiducial disk disruption model.
    \item SIDM subhalo abundances are suppressed relative to CDM at small Galactocentric radii.
    \item The suppression of subhalo abundance in WDM is roughly independent of Galactocentric radius.
\end{enumerate}

These results are consistent with simulations that directly include disk disruption (e.g., \citealt{Garrison-Kimmel170103792,Richings181112437,Webb200606695}) and with other SIDM simulations of MW-mass systems (e.g., \citealt{Robles190301469,Nadler200108754,Stafford200403872}). We emphasize that---in both the disk disruption and SIDM cases---the strength of subhalo disruption as a function of Galactocentric radius is model and resolution-dependent. For example, \cite{Green210301227} show that the abundance of surviving subhalos predicted by cosmological simulations within $\sim 0.1\ R_{\mathrm{vir}}$ is strongly resolution dependent due to both stripping below the mass resolution limit and artificial disruption. These numerical uncertainties may be exacerbated by imperfect halo finding in SIDM simulations \citep{Nadler200108754}.

Thus, SIDM and the disk preferentially suppress the abundance of subhalos in the inner regions of the MW, while WDM suppresses the abundance of subhalos that are expected to host dwarf galaxies in a radially-independent manner.

\subsection{LMC-associated Subhalos}
\label{sec:spatial}

We are uniquely poised to study the contribution of the LMC system to the total MW subhalo population because our simulations include realistic LMC analog systems. Averaged over our two CDM simulations, $60\pm 10\%$ of the total MW subhalo population with $V_{\mathrm{peak}}>20\ \mathrm{km\ s}^{-1}$ (including higher-order subhalos) is found in the hemisphere pointing toward the LMC from the host halo center, where the $1\sigma$ uncertainty includes both simulation-to-simulation and Poisson scatter, and is dominated by the latter. This anisotropy mainly results from the population of low-mass subhalos that fall into the MW with the LMC and is consistent with expectations from previous cosmological simulations (e.g., \citealt{Zentner0502496}). Referring to subhalos that were within the LMC's virial radius at the time of LMC infall as ``LMC-associated,'' we find that $\sim 80\%$ of the LMC-associated subhalos above our $V_{\mathrm{peak}}$ threshold are found in the LMC hemisphere at $z=0$, contributing $\sim 33\%$ to the total MW subhalo population in this direction. Recent discoveries and proper motion measurements of satellite galaxies in the southern hemisphere indicate similar anisotropy and LMC association statistics for the luminous counterparts of these subhalos (e.g., \citealt{PaperI,Patel200101746}).

\begin{deluxetable*}{{l@{\hspace{0.3in}}l@{\hspace{0.3in}}l@{\hspace{0.3in}}l@{\hspace{0.3in}}l@{\hspace{0.3in}}l@{\hspace{0.3in}}}}[t!]
\centering
\tablecolumns{5}
\tablecaption{Impact of SIDM, WDM, and the Galactic Disk on the MW Subhalo Population.}
\tablehead{
\colhead{Model\phantom{textttttt}} & \vtop{\hbox{\strut Mass-dependent}\hbox{\strut suppression}} & \vtop{\hbox{\strut Radially-dependent}\hbox{\strut suppression}} & \vtop{\hbox{\strut LMC-dependent}\hbox{\strut suppression}} & $f_{\mathrm{sub,LMC}}$ & Physical driver}
\startdata
SIDM
& \checkmark
& \checkmark 
& \checkmark 
& $48\pm 13\%$
& \vtop{\hbox{\strut Ram-pressure stripping}\hbox{\strut + tidal disruption}}
\\
WDM
& \checkmark
& \text{\sffamily X}$^*$
& \text{\sffamily X} 
& $58\pm 11\%$
& Free-streaming
\\
Disk
& \text{\sffamily X} 
& \checkmark 
& \text{\sffamily X} 
& $60\pm 10\%$
& Tidal disruption
\\
\enddata
{\footnotesize \tablecomments{The first column lists the dark matter or baryonic model. The second and third columns describe the suppression of the MW subhalo $V_{\mathrm{peak}}$ function and radial distribution, and the fourth column describes whether the suppression of LMC-associated subhalo abundances is enhanced relative to the suppression of the remaining subhalo population. The fifth column lists the fraction of $V_{\mathrm{peak}}>20\ \mathrm{km\ s}^{-1}$ subhalos in the hemisphere that points from the MW halo center toward the LMC, averaged over our two MW-like simulations. The final column describes the main physical effect that impacts the MW subhalo population. The asterisk indicates that the shape of the WDM radial distribution may change for subhalos below our fiducial $V_{\mathrm{peak}}$ threshold.}}
\label{tab:sidm_sims}
\end{deluxetable*}

The LMC-associated subhalo population and the spatial anisotropy it introduces in CDM respond differently to various mechanisms that suppress the formation or survival of low-mass subhalos. In the presence of baryons, ``preprocessing''---i.e., subhalo disruption before LMC infall into the MW within the LMC halo---may reduce the abundance of LMC-associated subhalos by up to $\sim 30\%$, which is significantly lower than the factor of $\sim 2$ disruption effect due to the MW disk \citep{Jahn190702979}. Because the efficiency of subhalo disruption due to the LMC galaxy is relatively uncertain and insignificant, our predictions only account for the impact of the MW disk on subhalo abundances. Thus, our Disk model does not preferentially disrupt LMC-associated subhalos because their radial distribution is approximately unbiased with respect to the rest of the MW subhalo population. In turn, anisotropy statistics are not significantly affected relative to CDM.

Our WDM model does not preferentially suppress the abundance of LMC-associated subhalos relative to the rest of the MW subhalo population. This is due to the fact that the $M_{\mathrm{peak}}$ distributions for MW and LMC-associated subhalos are similar. Thus, the spatial anisotropy introduced by the LMC system is slightly reduced in WDM but remains consistent with our CDM result within $1\sigma$ Poisson uncertainty.

SIDM has the most noticeable impact on the LMC-associated subhalo population, \emph{erasing} the anisotropy in our CDM simulations. In particular, $48\pm 13\%$ of subhalos with $V_{\mathrm{peak}}>20\ \mathrm{km\ s}^{-1}$ are found in the LMC hemisphere, averaged over our two SIDM simulations, where the $1\sigma$ uncertainty is again dominated by Poisson scatter. Moreover, the radial distribution of SIDM subhalos with respect to the LMC center is even more suppressed relative to the other models we consider compared to the radial distribution with respect to the MW center shown in Figure \ref{fig:vpeak_function}. Two effects reduce the abundance of LMC-associated subhalos and the resulting anisotropy of the MW subhalo population in SIDM:
\begin{enumerate}
    \item Preprocessing by the LMC: self-interactions with LMC halo particles disrupt $\sim 50\%$ of the LMC-associated subhalos above our $V_{\mathrm{peak}}$ threshold prior to LMC infall into the MW. These disrupted subhalos typically have maximum orbital velocities above $\sim200\ \mathrm{km\ s}^{-1}$; thus, this disruption is driven by the SIDM cross section at the same velocity scale that drives disruption within the MW.
    \item Ram-pressure stripping and tidal disruption by the MW halo: our LMC analogs are on high-velocity radial orbits into the MW, which enhances ram-pressure stripping due to self-interactions with MW halo particles, making LMC-associated subhalos even more susceptible to tidal disruption.\footnote{This mechanism also disrupts subhalos associated with the Gaia-Enceladus analogs in our simulations as they approach pericenter, amplifying the effect discussed in \cite{DSouzaMNRAS}.}
\end{enumerate}

Both of these effects depend on the specific history of the MW--LMC system; it is therefore crucial to study the impact of dark matter physics on subhalo populations constrained by the accretion history of the MW and the orbital properties of its most massive satellites. Although the anisotropy of the MW subhalo population in our SIDM simulations is only affected at the $\sim 1\sigma$ level due to the enhanced disruption of LMC-associated subhalos with $V_{\mathrm{peak}}>20\ \mathrm{km\ s}^{-1}$, Poisson uncertainties shrink quickly as this $V_{\mathrm{peak}}$ limit is decreased. Thus, probes of the MW and LMC subhalo populations below the galaxy formation threshold---including the statistics and density profiles of stellar streams over the full sky---will provide a powerful test of SIDM physics.

Interestingly, the clustering statistics of low-$V_{\mathrm{peak}}$ subhalos separated by $\lesssim 200\ \mathrm{kpc}$, quantified by the ratio of the subhalo--subhalo two-point correlation function relative to that in CDM, is significantly suppressed in our SIDM simulations, moderately suppressed in our Disk model, and not significantly affected in our WDM model. This is consistent with our finding that LMC-associated subhalos are preferentially suppressed in SIDM and indicates that the small-scale clustering properties of low-mass subhalos provides additional constraining power for such models. We leave a detailed study of this effect to future work.

Thus, LMC-associated subhalos are not preferentially affected by the Galactic disk or WDM relative to the rest of the MW subhalo population. Meanwhile, SIDM can enhance the disruption of LMC-associated subhalos due to self-interactions with the LMC at early times and during the LMC system's infall into the MW at late times, erasing the expected spatial anisotropy in the MW subhalo population and suppressing the small-scale clustering of low-mass subhalos relative to CDM.

\section{Discussion and Conclusion}
\label{sec:discussion}

We have demonstrated that the abundance and spatial distribution of the total MW subhalo population can be used to differentiate new kinds of dark matter physics from each other and from the effects of baryons due to the impact of the LMC system. These properties of the MW subhalo population have been inferred from observations of satellite galaxies and stellar streams, both of which are expected to rapidly advance in the coming years (e.g., \citealt{Drlica-Wagner190201055}). Furthermore, accurate predictions for the contribution of the LMC system to the MW subhalo population in alternative dark matter models will facilitate the interpretation of future direct \citep{Ibarra190800747} and indirect \citep{Ando190311427} detection experiments, which also rely on predictions for the abundance and spatial distribution of nearby subhalos.

Although we focused on subhalos with~$V_{\mathrm{peak}}>20\ \mathrm{km\ s}^{-1}$, which MW satellite observations currently probe, many of our qualitative results hold for the population of lower-mass subhalos probed by other tracers of substructure in the MW. For example, inferring the population of low-mass subhalos that perturb stellar streams is a promising avenue for studying dark matter physics (e.g., \citealt{Bonaca181103631,Banik191102662,Dalal201113141}). Our results imply that---in addition to dwarf galaxy abundances---the relative rate of subhalo--stream encounters toward and away from the LMC is sensitive to the SIDM cross section at the MW's infall velocity scale, which in turn informs a variety of dark matter particle models (e.g.\ \citealt{Tulin170502358,Correa200702958}).

Additional properties of the MW subhalo population are altered in each scenario we consider. For example, the density profiles of surviving subhalos in our SIDM simulations are often ``cored'' by self-interactions in a manner that depends on their infall time and orbital properties \citep{Nadler200108754}. This effect, which is not significant in viable WDM models \citep{Maccio12021282} and is less pronounced for low-mass subhalos in the presence of baryons \citep{Dutton201111351}, is also likely to be enhanced for LMC-associated subhalos in SIDM. Future spectroscopic measurements of MW and LMC-associated satellite galaxies (e.g., \citealt{Simon190304743}) can therefore further differentiate the impact of dark matter self-interactions from the effects of WDM-like models and baryons.

We note that \cite{Robles190301469} performed a related study, and we highlight several synergies and differences between this analysis and our work. \cite{Robles190301469} studied the subhalo $V_{\mathrm{max}}$ function and radial distribution of subhalos in a MW-mass host without an LMC analog system in CDM, SIDM, and disk disruption scenarios, finding that CDM and SIDM subhalo populations are similar in terms of their $V_{\mathrm{max}}$ functions and radial distributions in the presence of an embedded disk potential. The radial distributions in our SIDM simulations are consistent with these authors' SIDM-only findings, while our SIDM $V_{\mathrm{peak}}$ functions are more suppressed relative to CDM at low masses despite using a lower cross section, which likely results from including realistic LMC analog systems. We emphasize that, beyond the effects of the LMC, it is not clear whether different SIDM simulation codes yield consistent predictions for low-mass subhalo populations (e.g., \citealt{Nadler200108754}). Thus, future work that explores these potential discrepancies in detail is essential.

Our results indicate that the impact of dark matter physics on the MW subhalo population can depend on the specific history of the MW--LMC system. Future work that expands the available suite of MW-like simulations (D.\ Buch et al.\ 2021, in preparation) and explores modifications to zoom-in initial conditions to engineer MW-like mass accretion histories (e.g., \citealt{Roth150407250,Stopyra200601841,Rey210609729}) is therefore compelling.

\begin{acknowledgments}

We thank Tom Abel for useful discussions related to this work. 
This research received support from the National Science Foundation (NSF) under grant No.\ NSF DGE-1656518 through the NSF Graduate Research Fellowship received by E.O.N. 
This work received additional support from the Kavli Institute for Particle Astrophysics and Cosmology at Stanford at SLAC and from the U.S. Department of Energy under contract number DE-AC02-76SF00515 at SLAC National Accelerator Laboratory. Y.-Y.M.\ is supported by NASA through the NASA Hubble Fellowship grant No.\ HST-HF2-51441.001 awarded by the Space Telescope Science Institute, which is operated by the Association of Universities for Research in Astronomy, Inc., under NASA contract NAS5-26555. A.B.\ is supported by the Fermi Research Alliance, LLC under Contract No. DE-AC02-07CH11359 with
the U.S.\ Department of Energy (DOE), and the U.S.\ DOE Office of Science Distinguished Scientist Fellow Program.

This research made use of computational resources at SLAC National Accelerator Laboratory, a U.S.\ Department of Energy Office, and the Sherlock cluster at the Stanford Research Computing Center (SRCC); the authors are thankful for the support of the SLAC and SRCC computing teams.  This research made extensive use of \https{arXiv.org} and NASA's Astrophysics Data System for bibliographic information.

\end{acknowledgments}

\bibliographystyle{yahapj}
\bibliography{references}

\end{document}